\documentclass[11pt]{article}
\usepackage[T1]{fontenc}
\usepackage[utf8]{inputenc}
\usepackage{lmodern,microtype}
\usepackage[margin=1in]{geometry}
\usepackage{amsmath,amssymb,amsthm,mathtools}
\usepackage{booktabs,array,enumitem,needspace}
\IfFileExists{xurl.sty}{\usepackage{xurl}}{\usepackage[hyphens]{url}}
\usepackage[numbers,sort&compress]{natbib}
\usepackage{tikz}
\usetikzlibrary{arrows.meta,positioning,calc,fit,backgrounds}
\usepackage{xcolor,hyperref}
\hypersetup{colorlinks=true,linkcolor=blue!45!black,citecolor=blue!45!black,
 urlcolor=blue!45!black,pdftitle={Sharp Complexity Thresholds for Network Cournot Competition},
 pdfauthor={Mohammad Hajiaghayi and Mohammad Mahdavi}}
\newtheorem{theorem}{Theorem}[section]
\newtheorem{lemma}[theorem]{Lemma}
\newtheorem{proposition}[theorem]{Proposition}
\newtheorem{corollary}[theorem]{Corollary}
\theoremstyle{definition}

\newcommand{\R}{\mathbb R}
\newcommand{\Q}{\mathbb Q}
\newcommand{\Reg}{\operatorname{Reg}}
\newcommand{\poly}{\operatorname{poly}}
\newcommand{\PPAD}{\mathsf{PPAD}}
\newcommand{\one}{\mathbf 1}

\newcommand{\inputsize}{\mathcal L}
\setlist[itemize]{leftmargin=1.6em,itemsep=2pt,topsep=4pt}
\setlist[enumerate]{leftmargin=1.8em,itemsep=2pt,topsep=4pt}
\title{Sharp Complexity Thresholds for\\ Network Cournot Competition}
\author{Mohammad Hajiaghayi \qquad Mohammad Mahdavi\\[3pt]
  \normalsize University of Maryland, College Park\\
  \small\href{mailto:hajiagha@umd.edu}{\texttt{hajiagha@umd.edu}}
  \qquad\href{mailto:mahdavi@umd.edu}{\texttt{mahdavi@umd.edu}}}
\date{November 2026}
\begin{document}
\maketitle

\begin{abstract}
We determine the approximation complexity of network Cournot equilibrium
as a function of demand degree and the number of firms sharing a market.
For common, nonincreasing concave polynomial inverse demands,
convex quadratic costs of firms' allocation vectors, and rational
total capacities, high-precision approximation is polynomial-time
solvable for cubic demands, regardless of market size.
Quartic demands remain tractable when each market has at most three
firms, and quintic demands remain tractable with at most two. These
thresholds are exact: approximation at a fixed positive profit-regret
tolerance is PPAD-hard for
quartic demands with four firms per market, quintic demands with three,
and sextic demands with two. Each hardness result uses one fixed
demand curve in every market, a constant number of markets per firm,
and globally nondecreasing strongly convex quadratic costs.

We also give a polynomial-time algorithm for arbitrary-degree
nonincreasing concave polynomial demands when each firm's convex polynomial cost depends
only on its total production. The algorithm applies to arbitrary
access graphs and permits firm-specific inverse demands. Its running
time is polynomial in the input size and the logarithm of the inverse
accuracy. The algorithm assigns one marginal charge to each firm
and adjusts the charges until each firm's total sales approximate
the quantity it would produce at that charge.
\end{abstract}

\section{Introduction}

In network Cournot competition, a firm chooses how much to supply to
each market it can access. A market's price depends on its total
supply, while a firm's production cost couples its decisions across
markets. Two firms can compete in one region and face different
sets of competitors elsewhere.

Abolhassani, Bateni, Hajiaghayi, Mahini, and Sawant~\cite{ABHMS14}
developed an algorithmic treatment of this model. For affine inverse
demands, they gave a potential function that represents all firms'
incentives in a single convex optimization problem. Nonlinear demand
can destroy this potential structure. In that setting they used
complementarity methods under additional monotonicity hypotheses.

Two questions arise. First, how much nonlinearity is needed before
equilibrium computation becomes hard? Second, which forms of
production cost preserve efficient computation beyond the affine
case? Convexity alone does not answer either question. Each firm's
optimization problem is concave in its own quantities, but equilibrium
requires solving these problems simultaneously. Even strongly concave
polynomial games can be $\PPAD$-hard to approximate at constant
error~\cite{PVZ23}.
The issue is whether the specific common-price structure of Cournot
revenue prevents that hardness.

We answer the first question by giving an exact classification in
terms of two elementary parameters: the degree of the inverse-demand
polynomials and the maximum number of firms sharing a market. We
answer the second for costs of total production by a different
algorithm, which works for every explicitly encoded polynomial degree.
The distinction between total-production costs and general convex
quadratic allocation costs is essential. A cost of total production
couples markets only through one scalar; a quadratic cost of the full
allocation vector can couple different market quantities separately.

\subsection{Classification by demand degree and market size}

Let $r$ be a bound on demand degree and $h$ a bound on the number of
firms in each market. In this first result, every market has one
inverse-demand function common to its participating firms. Firms have
explicit rational total-production capacities and arbitrary convex
quadratic costs of their own allocation vectors. The demand
polynomials are nonincreasing and concave on the feasible intervals.
An additive $\varepsilon$ equilibrium is a feasible pure quantity
profile at which no firm can improve its profit by more than
$\varepsilon$ through a unilateral deviation.
A pure equilibrium exists throughout this class (Section~\ref{sec:model}).
A hard entry in Table~\ref{tab:classification} concerns the search
for an approximate equilibrium.
PPAD-hardness means that a polynomial-time algorithm for this search
would give polynomial-time algorithms for every problem in PPAD.

\begin{theorem}[Classification]
\label{thm:classification}
For the class above, the approximation complexity is given by
Table~\ref{tab:classification}. Each polynomial-time entry admits a
deterministic algorithm running in $\poly(\inputsize,b)$ time for
additive accuracy $2^{-b}$, where $\inputsize$ is the input length.
Each hard entry is $\PPAD$-hard for a universal constant positive
accuracy. Hardness already holds with one fixed identical demand
polynomial in all markets, a constant number of markets per firm,
and globally coordinatewise nondecreasing strongly convex quadratic
costs. The fixed-degree, bounded-local-dimension rational-output
versions defined in Appendix~\ref{sec:hard-membership} are
$\PPAD$-complete.
\end{theorem}

\begin{table}[t]
\centering
\renewcommand{\arraystretch}{1.13}
\begin{tabular}{@{}lcccc@{}}
\toprule
Demand degree bound & $h=1$ & $h=2$ & $h=3$ & $h\geq4$\\
\midrule
$r\leq3$ & P & P & P & P\\
$r=4$ & P & P & P & Hard\\
$r=5$ & P & P & Hard & Hard\\
$r\geq6$ & P & Hard & Hard & Hard\\
\bottomrule
\end{tabular}
\caption{The classification for common concave polynomial demands,
convex quadratic allocation costs, and rational total capacities.
P denotes polynomial time in input size and the logarithm of inverse accuracy;
Hard denotes PPAD-hardness at a fixed positive error. Degree and
participation are upper bounds on the input class.}
\label{tab:classification}
\end{table}

Thus the first hard degree is six when markets have at most two
firms, five when they have at most three, and four when four or more
firms are allowed. The classification is sharp unless $\PPAD$ search
admits polynomial-time algorithms. With one firm per market there are
no strategic interactions, and each firm's problem is concave
optimization for every degree.

In all three hard families, every market uses the same demand curve
and every firm's quadratic cost is strongly convex and nondecreasing
on the entire nonnegative orthant. Each firm has at most 26, 50, or 74
quantity coordinates in the sextic, quintic, or quartic construction,
respectively. All three constructions use the same fixed total
capacity, and prices are positive throughout the feasible region.

We obtain both sides of the classification from the same curvature
threshold. In Section~\ref{sec:classification} we bound how rapidly
the slope of a low-degree demand can change. Up to the market-size
cutoff, this bound makes the negative marginal-profit operator
monotone~\cite{Harks26}, so we can approximate an equilibrium by solving
a variational inequality~\cite{AFSZ26}. In Section~\ref{sec:hardness}
we choose a demand above the cutoff and link markets in a chain.
Along the chain we amplify one firm's influence on another while
keeping the reverse influence and the receiver's own payoff curvature
bounded. We use this asymmetry to encode a directed payoff interaction.

\subsection{Costs of total production}

The reduction couples individual deliveries through the firm's cost.
What if a firm pays the same production cost for any two allocations
with the same total output $S_i=\sum_jq_{ij}$? Under this restriction
we can compute an equilibrium at every demand degree.

\begin{theorem}[Costs of total production]
\label{thm:total-output}
Consider an arbitrary firm--market access graph and rational total
capacities. Suppose that firm $i$ has a convex polynomial cost
$C_i(S_i)$, and each inverse demand is a nonincreasing concave
polynomial on its feasible interval. Demands may be firm-specific.
For every $b\geq1$, a rational feasible additive $2^{-b}$ pure equilibrium can be
computed deterministically in $\poly(\inputsize,b)$ time.
Polynomial coefficients are rational and given by dense lists.
\end{theorem}

In Section~\ref{sec:total-output} we give each firm a marginal charge
for production and compute supply and sales separately at those charges.
Whenever a firm would produce more than it would sell, we can lower
its charge to bring those quantities closer together. To obtain a
polynomial bound on the number of updates, we control how much this
step widens the corresponding gaps for its rivals.

\subsection{Related work}
\label{sec:related}
\paragraph{Networked quantity competition.}
Abolhassani et al.~\cite{ABHMS14} introduced an algorithmic treatment of
Cournot competition on a bipartite firm--market graph. Their affine-demand
potential accommodates convex costs of each firm's market-specific
output vector; their nonlinear results use additional conditions on
the induced complementarity problem. Bimpikis et al.~\cite{BEI19}
study how the network determines quantities, profits, and the effects
of changes in market access. Their general model has common concave
demands and convex costs of total output. Their Proposition~1 states uniqueness for
strictly decreasing concave demands and convex increasing total costs;
their more explicit network formulas specialize the economic primitives.

Harks and Timmermans~\cite{HT22,HTcorr22} compute equilibria exactly in
polynomial time for rational atomic splittable singleton games with
player-specific affine costs. An equilibrium-preserving reduction
gives a multimarket Cournot application with firm-specific affine
inverse demands and quadratic costs of total production. Their separate work on nonlinear
convex-cost polymatroid games~\cite{HT18} gives a pseudo-polynomial
algorithm and a nonlinear Cournot application. Bhaskar and
Lolakapuri~\cite{BL18} study convex-cost atomic splittable routing and
give parallel-link algorithms whose running times are exponential in
the number of players or the number of links. Our total-cost theorem
allows both quantities to vary and gives a running time polynomial in
the input size and the logarithm of the inverse accuracy.
Hota and Kapoor~\cite{HK18} give strongly polynomial equilibrium
algorithms for a multiproduct Cournot model with substitutable
goods and quadratic utilities.

\paragraph{Marginal-profit formulations and capacities.}
Alsabah et al.~\cite{ABCIS21} analyze common concave inverse demands,
linear production and transport costs, and firm capacities. They
introduce a reduced game in marginal profits, prove uniqueness, and
establish global convergence of simultaneous best responses in those
variables. Caldentey and Haugh~\cite{CH24} characterize equilibria for
heterogeneous resource-constrained firms; their main computational
model has affine demand, and they also discuss nonlinear extensions.
Guo, Wang, and Candogan~\cite{GWC26} study production and shared
transportation capacities with endogenous congestion prices.
These models emphasize different ways in which resource constraints
couple markets. Our nonlinear algorithm treats independent firm
strategy sets and polynomial costs of total production, with an
explicit high-precision running-time bound.

Marginal-cost coordinates and contraction arguments have older
antecedents in Cournot and routing games~\cite{LS00,BL18}.
The sign structure of the reduced residual also relates to nonlinear
$Z$-function and $M$-function complementarity~\cite{Tamir74}.
To obtain a running time polynomial in the input size and the
requested precision, we bound the decrease in the total residual per
update and the accuracy needed in each scalar market solve.

\paragraph{Monotonicity and hardness.}
Harks~\cite{Harks26} gives curvature criteria for universal monotonicity
in atomic splittable congestion games. We specialize the market
criterion to common Cournot demand and identify which polynomial
degrees satisfy it for each market size. For those degrees we apply
the high-precision variational-inequality algorithm of
Anagnostides et al.~\cite{AFSZ26}.

Hardness for strongly concave polynomial games is already known:
Papadimitriou et al.~\cite{PVZ23} prove constant-error
$\mathsf{PPAD}$-hardness in that broader class. Our reductions impose
the specific Cournot revenue form, common inverse demand within a
market, convex nondecreasing quadratic production costs, and the small
market sizes at the classification boundary. They start from the
constant-error hardness of degree-three bipartite two-action
polymatrix games proved by Deligkas et al.~\cite{DFHM24}.
Klimm and Warode~\cite{KW25} establish $\mathsf{PPAD}$-completeness for
atomic splittable routing with player-specific affine costs on general
networks. Such routing strategy spaces differ from the singleton
structure underlying network Cournot games, so that result neither
gives our restricted hardness instances nor conflicts with our
positive results.

\section{Model and equilibrium approximation}
\label{sec:model}

An access graph is a bipartite graph $E\subseteq N\times M$, with
$n=|N|$ firms and $m=|M|$ markets. Write $E_i$ for the markets
accessible to firm $i$ and $N_j$ for the firms incident to market $j$.
Firm $i$ chooses a vector in
\[
 K_i=\left\{q_i\in\R_{\geq0}^{E_i}:
                   \sum_{j\in E_i}q_{ij}\leq D_i\right\},
 \qquad D_i\in\Q_{\geq0}.
\]
Let $K=\prod_iK_i$, $d=|E|$, $Q_j=\sum_{i\in N_j}q_{ij}$,
and $S_i=\sum_{j\in E_i}q_{ij}$. The common-demand model has profit
\begin{equation}
 \pi_i(q)=\sum_{j\in E_i}q_{ij}p_j(Q_j)-C_i(q_i).
 \label{eq:profit}
\end{equation}
Here $p_j$ is nonincreasing and concave on
$[0,B_j]$, where $B_j=\sum_{i\in N_j}D_i$, and $C_i$ is convex
on $K_i$. The polynomial-time algorithm for total-production costs
also permits firm-specific demands $p_{ij}$ and uses costs $C_i(S_i)$.

Each profit is concave in its own strategy: a market contribution
has own second derivative
$2p_j'(Q_j)+q_{ij}p_j''(Q_j)\leq0$, and the cost is convex.
The feasible sets are compact and convex. These games therefore
have pure equilibria by the standard existence theorem for concave
games~\cite{Rosen65}. Define actual unilateral regret by
\begin{equation}
 \Reg_i(q)=\max_{z_i\in K_i}
               \bigl(\pi_i(z_i,q_{-i})-\pi_i(q)\bigr).
 \label{eq:regret-definition}
\end{equation}
An additive $\varepsilon$ equilibrium has $\Reg_i(q)\leq\varepsilon$
for every firm. All approximation statements in this paper refer
to this profit guarantee.

Rational coefficients and capacities are represented in binary;
polynomial coefficient lists are dense. The input length
$\inputsize$ includes the graph and these descriptions. Accuracy
is specified as $2^{-b}$, and running times are measured in the
standard rational-input model. Arbitrary positive rational accuracy
is handled by a smaller dyadic target, including the supplied
accuracy's encoding in the input length. Convexity, monotonicity,
and concavity are promises on the specified domains. Zero-capacity
firms and isolated vertices are removed before running an algorithm.
Their deliveries are restored as zeros. If no coordinates remain,
the zero profile is an equilibrium.

For hardness, the reduction must decode any feasible profile that
satisfies the regret bound, including one with real-valued deliveries.
For PPAD membership we also need a finite output that can be checked
in polynomial time. In Appendix~\ref{sec:hard-membership} we specify
a polynomial encoding bound for rational outputs and verify the
regret condition when degree and local dimension are fixed.

\section{Polynomial-time regimes for allocation costs}
\label{sec:classification}

To compute equilibria in the polynomial-time regimes of
Table~\ref{tab:classification}, we show that the firms' negative
marginal-profit map is monotone. Consider a single market. An increase
in a rival's output can lower the price and steepen the demand curve.
The change in slope matters more to a firm with a large delivery
$q_i$, since its marginal revenue is $p(Q)+q_i p'(Q)$. We need a
bound that holds for every division of total output among the firms.

Write $t=-p'$. For $Q>0$ with $t(Q)>0$, measure the change in slope
by $Qt'(Q)/t(Q)$. We start by comparing the endpoint value of $t'$
with its average on $[0,Q]$, since
$t(Q)=t(0)+\int_0^Q t'(u)\,du$ and $t(0)\geq0$.
For a quartic demand, for instance, $t'$ is a nonnegative quadratic.
Its endpoint value is at most four times its average, so the ratio
is at most four. We next determine how large a ratio each market size
allows. With three suppliers the monotonicity cutoff is exactly
four~\cite{Harks26}, so the quartic bound suffices. For each degree
and market size that passes this test, we use monotonicity to compute
a profile with profit regret at most $2^{-b}$.

\begin{lemma}[Endpoint bounds]
\label{lem:upper-endpoint}
Let $f$ be a polynomial nonnegative on $[0,1]$. If its degree is at
most one, two, or three, respectively, then
\[
 f(1)\leq c\int_0^1 f(x)\,dx,
             \qquad c=2,\ 4,\ 6,
\]
respectively. Each constant is sharp.
\end{lemma}
\begin{proof}
For the three degree bounds, respectively, exact integration gives
\begin{align*}
 2\int_0^1 f(x)\,dx &= f(0)+f(1),\\
 4\int_0^1 f(x)\,dx &= 3f(1/3)+f(1),\\
 6\int_0^1 f(x)\,dx &= f(0)+4f(1/2)+f(1).
\end{align*}
Each identity follows by checking the monomials up to its stated
degree. All terms on the right are nonnegative, giving the desired
bounds on $f(1)$. Equality is attained by $x$, $(3x-1)^2$, and
$x(2x-1)^2$, respectively.
\end{proof}

\begin{corollary}[Curvature of low-degree demands]
\label{cor:upper-curvature}
Let $p$ be nonincreasing and concave on $[0,B]$. If its degree is at
most three, four, or five, then, wherever $Q>0$ and $-p'(Q)>0$,
\[
 0\leq\frac{Q(-p''(Q))}{-p'(Q)}\leq2,\ 4,\ 6,
\]
respectively.
\end{corollary}
\begin{proof}
Put $t=-p'\geq0$ and $f=t'\geq0$. If $f$ is zero the result
is immediate. Otherwise
$t(Q)=t(0)+\int_0^Qf(x)\,dx\geq\int_0^Qf(x)\,dx$.
Applying Lemma~\ref{lem:upper-endpoint} to $f(Qx)$ bounds
$Qf(Q)/t(Q)$ by the corresponding constant.
\end{proof}

To apply the curvature bound, consider a simultaneous change $d$ in
the firms' deliveries. If $J$ is the Jacobian of negative marginal
revenues, monotonicity requires $d^\top Jd\geq0$ along feasible
directions. Since $d^\top Jd=d^\top(J+J^\top)d/2$, it suffices to
check the symmetric part. Write $w_i=q_i/Q$ for firm $i$'s share
and $\kappa=Qt'(Q)/t(Q)$. Dividing the symmetric part by $t(Q)>0$
gives the matrix $S_h(w,\kappa)$ below.

\begin{lemma}[Market-size criterion]
\label{lem:upper-market}
For $h\geq2$ and $\kappa\geq0$, the matrix
\[
 S_h(w,\kappa)=I+\one\one^\top+
                       \frac\kappa2(w\one^\top+\one w^\top)
\]
is positive semidefinite for every probability vector
$w\in\R_{\geq0}^h$ if and only if
$\kappa\leq2(h+1)/(h-1)$.
\end{lemma}
\begin{proof}
The matrix depends affinely on $w$, and the positive semidefinite
cone is convex. It suffices to check the simplex vertices. By
symmetry take $w=e_1$. On vectors whose first coordinate and sum
of remaining coordinates are zero, the matrix is the identity.
An orthonormal basis for the remaining two directions is $e_1$ and
$(0,1,\ldots,1)/\sqrt{h-1}$. On this subspace the matrix is
\[
 \begin{pmatrix}
 2+\kappa &(1+\kappa/2)\sqrt{h-1}\\
 (1+\kappa/2)\sqrt{h-1}&h
 \end{pmatrix}.
\]
The diagonal entries are positive, and the determinant equals
\[
 \frac{h-1}{4}(\kappa+2)
                 \left(\frac{2(h+1)}{h-1}-\kappa\right).
\]
This proves both directions.
\end{proof}

We now compare the curvature bounds $2,4,6$ with the cutoff for each
market size. For the entries where the bounds meet the cutoff or fall
below it, we assemble the market contributions and convex costs into
a monotone variational inequality.

\begin{proposition}[Polynomial-time upper bounds]
\label{prop:upper-algorithm}
The P entries of Table~\ref{tab:classification} admit a rational
additive $2^{-b}$ equilibrium in $\poly(\inputsize,b)$ time, with
no restriction on local strategy dimension.
\end{proposition}
\begin{proof}
Let $G(q)=(-\nabla_{q_i}\pi_i(q))_i$. Fix a market with demand $p$,
total output $Q$, and $h$ firms, and put $t=-p'$. The negative
marginal revenue of firm $i$ is $-p(Q)+q_i t(Q)$. Its derivative
with respect to $q_\ell$ is $(1+\mathbf 1_{\{i=\ell\}})t(Q)+q_i t'(Q)$.
Thus, for $Q>0$ and $t(Q)>0$, the symmetric part of the market
Jacobian is
\[
 t(Q)S_h\!\left(w,\frac{Qt'(Q)}{t(Q)}\right),
 \qquad w_i=q_i/Q .
\]
The cutoff in Lemma~\ref{lem:upper-market} is $2+4/(h-1)$. It
equals six for two firms, four for three firms, and exceeds two for
every finite $h\geq2$. Thus the curvature bounds $6,4,2$ from
Corollary~\ref{cor:upper-curvature} give the quintic, quartic and
cubic entries, respectively. In particular, the quartic bound
includes equality for three firms. With four firms the cutoff is
$10/3$, so the bound of four no longer suffices. The lower bound for
that class is proved separately in Section~\ref{sec:hardness}.
With one firm, the contribution is the nonnegative scalar
$2t(Q)+Qt'(Q)$ for every degree. At $Q=0$, all deliveries
vanish and the unscaled matrix is $t(0)(I+\one\one^\top)$. If
$Q>0$ and $t(Q)=0$, the nonnegative, nondecreasing polynomial
$t$ vanishes on $[0,Q]$, so its derivative there is also zero. Convex costs
add positive semidefinite Hessians. Integrating along line segments
in $K$ proves that $G$ is monotone.

We apply the polynomial-time variational-inequality algorithm of
Anagnostides, Farina, Sandholm, and Zhang~\cite[Theorem~4.7]{AFSZ26}.
For a monotone continuous map on a compact convex domain, a
Stampacchia solution exists and is also a Minty solution. Their
algorithm therefore returns a feasible rational $q$ such that
\begin{equation}
                   G(q)^\top(z-q)\geq-2^{-b}
                         \qquad(z\in K).
 \label{eq:upper-vi}
\end{equation}
The following facts verify the computational assumptions. The map
$G$ is an explicit polynomial. Coefficient absolute-value bounds on
the capacity box bound its magnitude and Lipschitz constant by
numbers whose logarithms are polynomial in the input length.
Membership, separation, linear optimization, and Euclidean projection
over $K$ are rational polynomial-time operations.

If only one delivery coordinate survives preprocessing, a scalar
bisection suffices on its capacity interval. Own marginal profit is
nonincreasing. Its signs at the endpoints either certify an optimal
endpoint or bracket a zero. Given a coefficient bound $U$ on its
absolute value, reducing the bracket width to $2^{-b}/(1+U)$ bounds
the profit gap by $2^{-b}$. We may therefore suppose $d\geq2$ in
the ellipsoid construction.

The required well-bounded domain can be obtained by a rational
affine change of variables. After removing zero-capacity coordinates,
rescale $q_{ij}=D_i x_{ij}$. If firm $i$ has $d_i$ coordinates,
translate its unit-capacity simplex by the vector with all entries
$1/(d_i+1)$. Scale the product by $d(d+1)$. The transformed domain
contains the Euclidean unit ball and lies in a ball of radius
$d^2(d+1)$. Indeed, before scaling its inner radius is at least
$1/((d+1)\sqrt d)$ and its outer radius is at most $\sqrt d$.
Writing the inverse coordinate change as $q=Ay+b_0$, use
$\widetilde G(y)=A^\top G(Ay+b_0)$. This preserves the
variational-inequality gap exactly, as well as monotonicity, while
keeping all oracle descriptions and logarithmic bounds polynomial.
Euclidean projection onto the transformed product of simplexes is
also rational and polynomial-time: sort coordinates and subtract a
common threshold in each block.
These are the well-bounded-domain hypotheses used in~\cite{AFSZ26}.

Finally, in~\eqref{eq:upper-vi} let $z$ differ from $q$ only in
firm $i$'s coordinates. Concavity of that firm's profit gives
\[
 \pi_i(z_i,q_{-i})-\pi_i(q)
 \leq\nabla_{q_i}\pi_i(q)^\top(z_i-q_i)\leq2^{-b}.
\]
Thus the returned solution has the required actual regret.
\end{proof}

\section{PPAD-hardness at the degree thresholds}
\label{sec:hardness}

At the first degree beyond each polynomial-time regime, we use the
extra curvature to encode a two-action polymatrix game in a Cournot
economy. Each source player becomes a firm whose deliveries to two
private markets encode its action probabilities. We need to recover a
$1/20$-equilibrium of the source game from \emph{any} sufficiently
accurate equilibrium of the economy. To do so, we arrange that a
source deviation worth more than $1/20$ gives the corresponding firm
a unilateral deviation worth more than the Cournot regret tolerance.

Consider a payoff entry that contributes $-Lxy$ to the receiver when
the receiver and sender play the corresponding actions with probabilities
$x$ and $y$. Holding $y$ fixed, an increase of $\Delta x$ costs the
receiver $Ly\,\Delta x$, regardless of its initial $x$. On a single
shared market the sender can affect the price, but the receiver's
marginal revenue also falls as its own output rises. An interior
quantity can therefore be optimal even when the source player strictly
prefers a pure action. We insert relay firms between the sender and
receiver to amplify the term proportional to $y$ while keeping the
receiver's own payoff curvature and the influence in the opposite
direction bounded. Each shared market has one high-output and $s$
low-output firms. For $s=1,2,3$ we use degrees $6,5,4$, respectively.

\begin{theorem}[Fixed-demand hardness]
\label{thm:hardness}
There is a universal rational $\varepsilon_0>0$ with the following
property.  For each $(r,h)\in\{(6,2),(5,3),(4,4)\}$ there is one fixed
rational polynomial $p_r$, nonincreasing and concave on
$\mathbb R_{\geq0}$, for which computing an additive
$\varepsilon_0$ pure Nash equilibrium is PPAD-hard.  Every market uses
$p_r$, every firm has total capacity $1000$, and prices are positive on
the feasible quantity range.  Costs are strongly convex quadratics,
are coordinatewise nondecreasing on the entire nonnegative orthant,
and vanish at zero.
There are at most $h$ firms per market and at most $24(h-1)+2$ markets
per firm.  The bounded rational-output search relations specified in
Appendix~\ref{sec:hard-membership} are PPAD-complete.
\end{theorem}

We build the reduction in two stages. First we work in an auxiliary
quadratic game, where we can solve for the relay signals at fixed
endpoint strategies. When we then measure a player's gain from
deviating, we hold those signals fixed because only that player can
move. Second we implement the quadratic game with
Cournot deliveries. We use costs to keep each firm's deliveries
consistent with a single strategy, and we choose the delivery scale
so that the quadratic terms determine its incentives up to a small error.

\subsection{Two-action polymatrix games}

Take as input a bipartite polymatrix game of maximum degree three,
with two actions per player and total payoffs normalized to $[0,1]$.
Computing a $1/20$-Nash equilibrium for these games is PPAD-hard
by Deligkas, Fearnley, Hollender, and Melissourgos~\cite[Theorem~4.4]{DFHM24}.
To assign one relay chain to each payoff entry, we express each
player's payoff as a sum of losses with coefficients in $[0,1]$.
The losses must preserve the payoff difference between the player's
two actions against any fixed profile of its neighbors. Write $x_{ia}$
for the probability of action $a$. We seek directed loss matrices
$L^{ij}\in[0,1]^{2\times2}$ giving payoffs of the form
\begin{equation}
 u_i^{\rm loss}(x)=-\sum_{j\in N(i)}\sum_{a,b\in\{0,1\}}
                       L^{ij}_{ab}x_{ia}x_{jb}.
 \label{eq:hard-loss-source}
\end{equation}
To choose the entries, write player $i$'s action-one minus
action-zero payoff difference as
$c_i+\sum_j d_{ij}x_{j1}\in[-1,1]$.
For each nonisolated $i$, choose $c_{ij}$ summing to $c_i$ in the intervals
\[
 -1-\min\{d_{ij},0\}\leq c_{ij}\leq1-\max\{d_{ij},0\}.
\]
These intervals are nonempty because $|d_{ij}|\leq2$.  Their summed
endpoints contain $c_i$, by minimizing and maximizing the payoff
difference over the neighbor cube.  A greedy interval allocation
therefore finds rational $c_{ij}$ in polynomial time.  Set
$D_{ij}(b)=c_{ij}+d_{ij}b$,
$L^{ij}_{0b}=(D_{ij}(b))_+$, and
$L^{ij}_{1b}=(-D_{ij}(b))_+$.
Then $L^{ij}_{0b}-L^{ij}_{1b}=D_{ij}(b)$. Summing over neighbors
gives the original difference between pure actions for every profile
of the opponents. Affinity in the player's own strategy gives equality
of payoff differences for mixed deviations as well.
Isolated players can be restored after decoding.

\subsection{Equilibria of the quadratic relay game}
\label{sec:hard-relays}

Fix the strategies at the two ends of one chain. To find what signal
reaches the receiver, give each relay a baseline with a large delivery
on its incoming market and small deliveries on its outgoing markets.
Its signal specifies a change from these quantities. A relay responds
to its neighbors on both sides, so we solve for all signals
simultaneously. Since we must decode approximate equilibria, we also
bound how far the signals can stray from this solution when each
relay has small regret.

Let $J$ be a fixed polynomial satisfying
$J(0)=J'(0)=0$, $J'(1)=1$, $0<J(1)<1$, and
$J''(1)=\kappa>0$.
Use the inverse demand $p(Q)=A-J(Q)$. We will choose the price level
$A$ when we specify the costs.
Choose $0<\beta<1/(s+1)$ and $0<\vartheta<2$, and baseline quantities
\[
 \beta_R=1-s\beta,\qquad \beta_L=\beta,
\]
so that one high-output firm and $s$ low-output firms supply total
quantity one.  A high firm's signal changes its quantity by
$-\delta\vartheta z$, and a low firm's signal changes its quantity by
$\delta z$.  Define
\begin{align}
 a&=1+\kappa\beta_R,& b&=1+\kappa\beta,&
 R_0&=a+1,& S_0&=b+1,\notag\\
 d_0&=R_0\vartheta^2+sS_0,&
 \alpha&=R_0\vartheta^2+s,& \gamma&=sb,\notag\\
 D&=\alpha+s\gamma,& A_0&=sa\vartheta,& B_0&=sb\vartheta.
 \label{eq:relay-coefficients}
\end{align}
At the limiting baseline $\beta=0$, the expression
$A_0+B_0-D=(\kappa+2)(s\vartheta-\vartheta^2)-s-s^2$
is maximized at $\vartheta=s/2$. Substitution gives
\[
 A_0+B_0-D=\frac{s}{4}(s\kappa-2s-4).
\]
Thus at this limiting choice $A_0+B_0>D$ holds exactly when
$\kappa>2+4/s$. Since a shared market has $h=s+1$ firms, the
right-hand side is the monotonicity cutoff from Lemma~\ref{lem:upper-market}.
If also $A_0>B_0$ and $D>2\sqrt{A_0B_0}$, this
inequality puts $A_0$ above the larger root of
$x^2-Dx+A_0B_0$, producing a growing forward transfer along the chain.
Our positive baseline $\beta$ preserves this strict gap.
The constants will satisfy
\begin{equation}
 \begin{gathered}
 d_0<20,\quad\alpha>3,\quad D-2\sqrt{A_0B_0}>3/2,
 \quad (A_0/2+2B_0)/D<5/6,\\
 A_0<23,\quad B_0<5,\quad A_0B_0<102,
 \quad B_0/A_0<1/3,\quad a/b<10,\\
 R_0\vartheta^2<14,\qquad sS_0<7.
 \end{gathered}
 \label{eq:relay-required-bounds}
\end{equation}

For an integer $k\geq2$, let $T_k$ be the tridiagonal matrix with diagonal $D$, subdiagonal
$-A_0$, and superdiagonal $-B_0$.  With
$\theta_j=\det T_j$ and $\theta_0=1$, put
\begin{equation}
 K=\frac{A_0^{k+1}}{\theta_k},\qquad
 K_- =\frac{B_0^{k+1}}{\theta_k},\qquad
 E=\frac{A_0B_0\theta_{k-1}}{\theta_k}.
 \label{eq:relay-gains}
\end{equation}
We choose a fixed chain length $k$ in Section~\ref{sec:constants}
to satisfy
\begin{equation}
 2^{41}<K<2^{k+8},\qquad K_-<1,\qquad 0<E<100.
 \label{eq:relay-gain-bounds}
\end{equation}
Thus $K$ is a large forward gain, $K_-$ a small reverse gain, and $E$
a bounded feedback coefficient.

There are $s$ scalar relay players in each of $k$ layers.  Their strategy
interval is $[-H,H]$, where $H=2^{k+4}$.
Let $\mathcal M$ have diagonal blocks
$\alpha I_s+\gamma\mathbf1\mathbf1^\top$, subdiagonal blocks
$-a\vartheta\mathbf1\mathbf1^\top$, and superdiagonal blocks
$-b\vartheta\mathbf1\mathbf1^\top$.  Each own diagonal entry is $d_0$.
For endpoint signals $u,v\in[-1/2,1/2]$ and a loss $L\in[0,1]$,
the vector of derivatives of the relays' payoffs with respect to
their own signals is $f-\mathcal Mz$, where $f$ is uniform
within the first and last layers.  The first-layer entry is
$(A_0/s)u$, the last-layer entry is $-B_0Lv$, and all others are zero.

\begin{lemma}[Stable amplification]
\label{lem:hard-relay}
The unique unconstrained relay solution is uniform in each layer,
$z^*_{\ell h}=y^*_\ell$, lies in the interior of $[-H,H]^{sk}$, and
satisfies
\begin{equation}
 \begin{split}
 T_ky^*&=(A_0/s)u e_1-B_0Lv e_k,\\
 y_k^*&=\frac{K}{sA_0}u-\frac{E}{A_0}Lv,
 \qquad
 y_1^*=\frac{E}{sB_0}u-\frac{K_-}{B_0}Lv.
 \end{split}
 \label{eq:relay-boundary-identities}
\end{equation}
If each relay has regret at most $\xi$, then
\begin{equation}
 \|z-z^*\|_\infty\leq\sqrt{60k\,10^k\xi}.
 \label{eq:relay-approximation}
\end{equation}
The parameters will make this bound less than $1/1000$.
\end{lemma}
\begin{proof}
In the weighted maximum norm
$\|y\|_w=\max_\ell |y_\ell|/2^\ell$, the nonnegative matrix
$I-T_k/D$ has norm below $5/6$.  Hence $T_k$ is invertible, its inverse
is nonnegative, and
$\|T_k^{-1}\|_w<6/D$.
The forcing in the first line of
\eqref{eq:relay-boundary-identities} has weighted norm
at most $\max\{A_0/(4s),B_0/2^{k+1}\}$.
The resulting solution obeys $|y^*_\ell|<4\cdot2^\ell\leq H/4$.
Tridiagonal cofactor expansion gives
\[
 (T_k^{-1})_{k1}=A_0^{k-1}/\theta_k,\quad
 (T_k^{-1})_{1k}=B_0^{k-1}/\theta_k,\quad
 (T_k^{-1})_{11}=(T_k^{-1})_{kk}=\theta_{k-1}/\theta_k,
\]
which proves the boundary identities.

For stability, use layer weights $\omega_\ell=(b/a)^{\ell-1}$ and
$W=\operatorname{diag}(\omega_1I_s,\ldots,\omega_kI_s)$.
The layer weights balance each pair of opposite coefficients, so
$W\mathcal M$ is symmetric. To bound its quadratic form relative
to the weighted norm, put $S=W^{1/2}\mathcal M W^{-1/2}$. This
symmetric matrix has eigenvalue $\alpha$ on each within-layer
difference subspace. On the uniform-layer subspace it has diagonal
$D$ and off-diagonal $-\sqrt{A_0B_0}$. Its least eigenvalue there
is at least $D-2\sqrt{A_0B_0}>3/2$. Applying these bounds to
$W^{1/2}z$ gives
\[
 z^\top W\mathcal Mz\geq\tfrac32\|z\|_W^2,
 \qquad \min_\ell\omega_\ell>10^{-k}.
\]
This also proves uniqueness of the full linear solution and its
uniformity within layers.

Put $\Delta=z-z^*$.  Relay $i$ may unilaterally move one tenth of the
way toward $z_i^*$.  Its improvement is
\[
 \tfrac1{10}\Delta_i(\mathcal M\Delta)_i
       -\tfrac{d_0}{200}\Delta_i^2.
\]
The move is feasible even if $z_i$ is on the boundary.  Multiply these
individual regret inequalities by $W_i$ and sum.  Since $s\leq3$,
\[
 3k\xi\geq\sum_iW_i\xi
 \geq\left(\frac3{20}-\frac{20}{200}\right)\|\Delta\|_W^2
 =\frac1{20}\|\Delta\|_W^2.
\]
The lower bound on the weights proves \eqref{eq:relay-approximation}.
In particular, this argument neither assumes interiority of the
approximate profile nor permits a coordinated deviation.
\end{proof}

\subsection{Market construction and source incentives}

We now construct markets with the quadratic incentives just specified.
With two firms per market, consecutive relays meet in pairs. One
sells a small quantity on a market where the next sells a large
quantity. A change in the price slope has a larger effect on the
large seller's marginal revenue, giving the link its direction.
With $s+1$ firms, the next relay responds to the combined deliveries
of $s$ predecessors.
All these markets use the same demand curve. To encode a particular
loss $L$, we vary the receiver's delivery with $L$ and add a linear
cost. We can then isolate the multiple of $L$ in the receiver's
marginal payoff and bound the other terms across its incident chains.

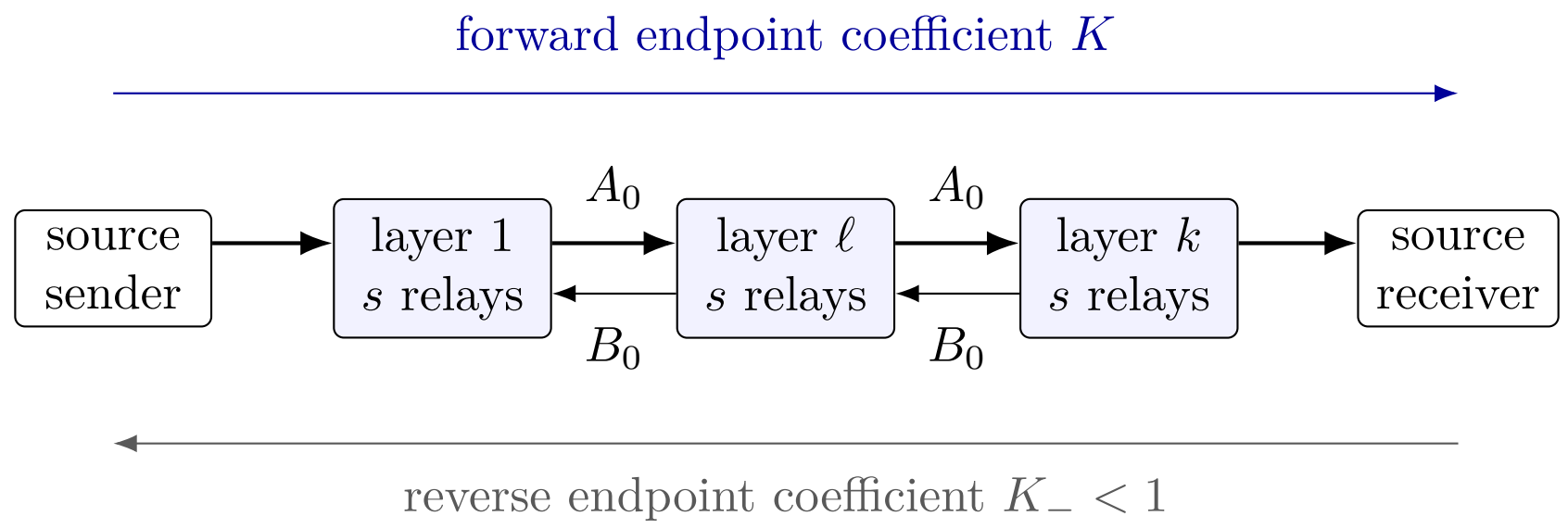
\begin{figure}[t]
\centering
\begin{tikzpicture}[x=1cm,y=1cm,>=Latex,
 block/.style={draw,rounded corners=2pt,minimum width=1.55cm,
 minimum height=.8cm,align=center,font=\small},
 endpoint/.style={draw,rounded corners=2pt,minimum width=1.4cm,
 minimum height=.8cm,align=center,font=\small}]
 \node[endpoint] (u) at (0,0) {source\\sender};
 \node[block,fill=blue!5] (l1) at (2.35,0) {layer 1\\$s$ relays};
 \node[block,fill=blue!5] (lm) at (4.8,0) {layer $\ell$\\$s$ relays};
 \node[block,fill=blue!5] (lk) at (7.25,0) {layer $k$\\$s$ relays};
 \node[endpoint] (v) at (9.6,0) {source\\receiver};
 \draw[->,thick] ($(u.east)+(0,.18)$)--($(l1.west)+(0,.18)$);
 \draw[->,thick] ($(l1.east)+(0,.18)$)--node[above=3pt,font=\small]{$A_0$}($(lm.west)+(0,.18)$);
 \draw[->,thick] ($(lm.east)+(0,.18)$)--node[above=3pt,font=\small]{$A_0$}($(lk.west)+(0,.18)$);
 \draw[->,thick] ($(lk.east)+(0,.18)$)--($(v.west)+(0,.18)$);
 \draw[->] ($(lm.west)+(0,-.18)$)--node[below=3pt,font=\small]{$B_0$}($(l1.east)+(0,-.18)$);
 \draw[->] ($(lk.west)+(0,-.18)$)--node[below=3pt,font=\small]{$B_0$}($(lm.east)+(0,-.18)$);
 \draw[->,blue!60!black] (0,1.25)--node[above=3pt,font=\small]{forward endpoint coefficient $K$}(9.6,1.25);
 \draw[->,gray!70!black] (9.6,-1.25)--node[below=3pt,font=\small]{reverse endpoint coefficient $K_-<1$}(0,-1.25);
\end{tikzpicture}
\caption{The relay construction, shown schematically. Each block
represents a layer of one, two, or three firms. Each internal arrow
represents a typical adjacent-layer interaction; intervening layers
are suppressed. A common Cournot market creates unequal local cross
effects. Repetition amplifies forward influence while keeping the
reverse endpoint effect bounded.}
\label{fig:relays}
\end{figure}

For each directed interaction $i\leftarrow j$ and action pair $(a,b)$,
make a fresh relay chain and set
\[
 L=L^{ij}_{ab},\qquad u=x_{jb}-1/2,\qquad v=x_{ia}-1/2.
\]
For every relay $h$ of layer $\ell\geq2$, create one market whose
high firm is that relay and whose $s$ low firms are the previous
layer.  Their canonical quantities are
\begin{equation}
 q_R=\beta_R-\delta\vartheta z_{\ell h},\qquad
 q_{L,t}=\beta+\delta z_{\ell-1,t}.
 \label{eq:relay-market-quantities}
\end{equation}
At the first layer, one low firm is source firm $j$, supplying
$\beta+\delta u$; the other $s-1$ low firms are fresh dummies,
each with canonical quantity $\beta$.  This creates $s$ first-layer
markets and never places a firm in two roles in one market.
At the last boundary, make $s$ separate markets whose high firm is
source firm $i$ and whose low firms are all last-layer relays, with
\begin{equation}
 q_R=\beta_R+\delta\vartheta Lv,\qquad
 q_{L,t}=\beta+\delta z_{kt}.
 \label{eq:relay-last-market}
\end{equation}
Every relay is high in one market and low in $s$ markets. Figure~\ref{fig:relays} shows the incidence between layers.

Write $\beta_e$ for a firm's baseline delivery in a shared market.
For raw demand $-J$, subtract the linear cost $\ell_e q_e$ with
\begin{equation}
                  \ell_e=-J(1)-\beta_e.
 \label{eq:relay-linear-cancellation}
\end{equation}
Since $J'(1)=1$, this cancels the own first derivative of
$-q_eJ(Q)$ at the baseline. The own second derivative is
$-2-\kappa\beta_e$ and each cross derivative of own marginal
revenue is $-1-\kappa\beta_e$. Consider relay $h$ in an interior
layer $\ell$. It is high in one market, with signal coefficient
$-\vartheta$, and low in $s$ markets, with coefficient $1$, in
units of $\delta$. Its normalized quadratic payoff has derivative
\[
 -d_0 z_{\ell h}
 -sb\sum_{t\ne h}z_{\ell t}
 +a\vartheta\sum_t z_{\ell-1,t}
 +b\vartheta\sum_t z_{\ell+1,t}.
\]
Indeed, its own coefficient is $-R_0\vartheta^2-sS_0=-d_0$.
Each other same-layer relay shares its $s$ outgoing markets.
At the first boundary the preceding-layer sum is $u$, and at the
last boundary the next-layer sum is $-sLv$. Since $d_0=\alpha+\gamma$
and $\gamma=sb$, these derivatives are the entries of
$-\mathcal Mz+f$ in Lemma~\ref{lem:hard-relay}.

Let $\bar z_\ell=s^{-1}\sum_hz_{\ell h}$.
Add a linear source cost $(K/2)Lx_{ia}$ in the normalized quadratic
game.  For one chain the source receiver's quadratic contribution is
\[
 -\frac{sR_0\vartheta^2L^2}{2}v^2-sA_0Lv\bar z_k-(K/2)Lx_{ia},
\]
and the source sender's is
\[
                         -\frac{sS_0}{2}u^2+B_0u\bar z_1.
\]
The normalized quadratic game sums these endpoint payoffs over
incident chains. Each source firm chooses the pair
$(x_{i0},x_{i1})$ in the probability simplex, and each relay chooses
its own coordinate in $[-H,H]$.

\begin{lemma}[Source regret in the quadratic game]
\label{lem:quadratic-source-regret}
Assume \eqref{eq:relay-required-bounds} and \eqref{eq:relay-gain-bounds}.
Let $0<\xi\leq1$ satisfy $\sqrt{60k\,10^k\xi}<1/1000$.
At any $\xi$-equilibrium of the normalized quadratic game,
the source profile satisfies
\begin{equation}
 \operatorname{Reg}^{\rm source}_i(x)
       \leq\frac{\xi+5000}{K}<1/20
       \qquad\text{for every source player }i .
 \label{eq:relay-source-decoding}
\end{equation}
\end{lemma}
\begin{proof}
Fix the profile and differentiate each endpoint payoff with the relays
held fixed. Substituting~\eqref{eq:relay-boundary-identities} at this
profile gives the receiver and sender derivatives, respectively,
\begin{equation}
 \begin{split}
 &-KLx_{jb}+s(E-R_0\vartheta^2)L^2v
                   -sA_0L(\bar z_k-y_k^*),\\
 &(E/s-sS_0)u-K_-Lv
                   +B_0(\bar z_1-y_1^*).
 \end{split}
 \label{eq:relay-endpoint-gradients}
\end{equation}
For one incoming chain, the difference from the target derivative
$-KLx_{jb}$ is at most $3(100+14)/2+3\cdot23/1000<172$.
For one outgoing chain the absolute derivative is at most
$(100+7)/2+1/2+5/1000<55$. Here we used $|u|,|v|\leq1/2$,
$0\leq L\leq1$, and the relay error bound of Lemma~\ref{lem:hard-relay}.
Each source action occurs in at most six chains in each direction,
so its derivative differs from $K$ times the derivative of
\eqref{eq:hard-loss-source} by at most $6(172+55)<2000$.

With all relays held fixed, a source firm's own quadratic Hessian is
diagonal and has entries of absolute value below
$6sR_0\vartheta^2+6sS_0<300$.
A best source deviation can be taken pure because the source payoff
is affine in the player's own probability. A deviation to a pure
source action has $\ell_1$ distance at most two
and squared $\ell_2$ distance at most two. Hence the gradient error
costs at most $2000\cdot2$ in the gain from this move. The firm's
own quadratic term costs at most $\tfrac12\cdot300\cdot2$ more.
Taking $g=\operatorname{Reg}^{\rm source}_i(x)$, this feasible
unilateral move gains at least $Kg-4300$ and at most $\xi$.
Equation~\eqref{eq:relay-source-decoding} follows from $K>2^{41}$.
\end{proof}

\subsection{Quadratic costs and consistency of deliveries}
\label{sec:hard-implementation}

The quadratic game gives a source player one mixing probability for
all its neighbors. In the Cournot economy the corresponding firm can
choose a different delivery on every market. To make these choices
encode one strategy, we record it in the firm's private deliveries
and penalize discrepancies between that record and its shared deliveries.
We design the penalty so that the firm can eliminate it unilaterally
by choosing any consistent strategy, including either pure action.

Each non-dummy firm serves two private markets. The canonical private
deliveries of a source firm are $r_{ia}=\delta x_{ia}$, and those
of a relay are $r_{i0}=\delta(z_i+H)$ and
$r_{i1}=\delta(H-z_i)$.
The required signal sums are $D_i^{\rm sig}=1$ and $2H$, respectively.
For such a firm each shared canonical quantity has the form $\beta_e+\delta d z$,
where $z=x_{ia}-1/2$ for a source action and $z=z_i$ for a relay.
Put $T=1/2$ or $T=H$, respectively. The firm's two private coordinates
are $\delta(T+z)$ and $\delta(T-z)$, up to ordering. For $d\geq0$
use $r=\delta(T-z)$, and for $d<0$ use $r=\delta(T+z)$. In both cases
$\beta_e+\delta d z=(\beta_e+\delta |d|T)-|d|r$.
Since $|d|\leq\max\{1,\vartheta\}<2$ and $2\delta H<\beta/2$,
the resulting rows satisfy
\begin{equation}
 q_{ie}=h_{ie}-c_{ie}r_{i,a(i,e)},
       \qquad 0\leq c_{ie}<2,\quad 0<h_{ie}<2.
 \label{eq:positive-copy-rows}
\end{equation}

A canonical strategy consists of the private pair for an admissible
source probability or relay signal and all its shared deliveries
given by~\eqref{eq:positive-copy-rows}. A dummy's canonical strategy
is the singleton $\{\beta\}$. Thus choosing a new signal changes only
the corresponding firm's delivery vector. The capacity bound below
makes each such unilateral deviation feasible.

For non-dummy firms, set
\begin{equation}
 \Phi_i=
 \left((r_{i0}+r_{i1})/\delta-D_i^{\rm sig}\right)^2
 +\sum_e(q_{ie}+c_{ie}r_{i,a(i,e)}-h_{ie})^2.
 \label{eq:relay-penalty}
\end{equation}
A dummy has one coordinate and penalty $(q_i-\beta)^2$.
Give source private coordinate $a$ the nonnegative linear coefficient
\begin{equation}
            v_{ia}=\frac{\delta K}{2}\sum_{j,b}L^{ij}_{ab}.
 \label{eq:relay-source-bias}
\end{equation}
Other private coefficients are zero.  Our parameters ensure
$v_{ia}\leq3$.
Every market, including each private market, uses $p(Q)=A-J(Q)$,
and the cost is
\begin{equation}
 C_i(q_i)=A\sum_eq_{ie}
       +\sum_{e\text{ shared}}\ell_eq_{ie}
       +\sum_av_{ia}r_{ia}
       +M\bigl(\Phi_i(q_i)-\Phi_i(0)\bigr).
 \label{eq:implemented-vector-cost}
\end{equation}
The common affine shift cancels from the payoff.
Because the rows in \eqref{eq:relay-penalty} have nonnegative
coefficients, the quadratic Hessian is both positive semidefinite
and entrywise nonnegative.  A shared cost partial at zero is at least
$A-2-4M$.  A private partial is at least
$A-4MH/\delta-576M$: there are at most $72$ shared coordinates and
each row has coefficient and intercept below two.
Choosing
\begin{equation}
                 A>2+4MH/\delta+576M
 \label{eq:shift-requirement}
\end{equation}
makes every cost globally coordinatewise nondecreasing.
The dummy partial has the same shared-coordinate bound.
The subtraction of $M\Phi_i(0)$ gives $C_i(0)=0$.

All canonical shared quantities are strictly positive, since
$2\delta H<\beta/2$. A private coordinate can be zero, for example
when its action has probability zero. Decoding uses the sum of the
private pair, which is positive. A source uses at most $24s$ shared markets
and two private markets.  A relay uses $s+1$ shared and two private
markets; a dummy uses one market.  Their canonical total quantities
are below $145$, hence feasible under capacity $1000$.
A source has at most $12s(s+1)$ distinct opponents: an outgoing
chain has $s^2$ first-layer opponents, an incoming chain has $s$
last-layer opponents, and there are at most twelve chains of each kind.

\begin{lemma}[Decoding arbitrary approximate profiles]
\label{lem:hard-decoding}
Suppose raw payoffs before penalties and the cancelling shift have
absolute value at most $R$ on the capacity box, and every partial
derivative with respect to an own coordinate or an incident opponent
coordinate has absolute value at most $L_0$.
Let $M=3R/\rho^2$, $0<\rho<1$, and $0<\varepsilon_*<R$.
Every actual $\varepsilon_*$-equilibrium decodes in polynomial time
to a canonical profile with raw canonical regret at most
\begin{equation}
                         \varepsilon_*+1600L_0\rho.
 \label{eq:decoded-raw-regret}
\end{equation}
\end{lemma}
\begin{proof}
Comparison with any canonical zero-penalty strategy gives
$M\Phi_i\leq2R+\varepsilon_*<3R$; hence $\Phi_i\leq\rho^2$.
For a non-dummy firm, normalize its private pair to the required sum:
\[
 \widehat r_{ia}
   =\frac{\delta D_i^{\rm sig}r_{ia}}{r_{i0}+r_{i1}}.
\]
The denominator is positive because its distance from
$\delta D_i^{\rm sig}$ is at most $\delta\rho<\delta D_i^{\rm sig}$.
The private $\ell_1$ error is at most $\delta\rho$ and each shared
coordinate error is at most $\rho+2\delta\rho\leq3\rho$.
A dummy is decoded to $\beta$, with error at most $\rho$.

The total own-coordinate error is at most $217\rho$.
Each of at most $72$ incident markets has at most three opponents,
so the relevant opponent-coordinate error is at most $648\rho$.
For a canonical unilateral deviation, the actual equilibrium
inequality bounds its raw gain by
$\varepsilon_*-M\Phi_i(q_i)\leq\varepsilon_*$.
Replacing the current profile by its canonical decoding changes
the current raw payoff by at most $(217+648)L_0\rho$, and replacing
only the opponents in the deviating payoff changes that payoff by
at most $648L_0\rho$.  Their sum is less than $1600L_0\rho$.
All comparisons remain inside the capacity box.
For dummies the canonical strategy set is a singleton.
\end{proof}

\subsection{Regret transfer and completion of the reduction}

Lemma~\ref{lem:hard-decoding} bounds the gain from changing a
canonical delivery vector. To apply the source-regret bound, we
must express that gain in the units of the quadratic game. The
quadratic terms scale as $\delta^2$, and for a fixed chain the Taylor
remainder scales as $\delta^3$. We choose $\delta$ small enough that
the error after division by $\delta^2$ fits within the relay tolerance.

\begin{proof}[Proof of Theorem~\ref{thm:hardness}]
First omit the strongly convex perturbation below. Let $q$ be any
$\varepsilon_*$-equilibrium and let $\widehat q$ be the canonical
profile given by Lemma~\ref{lem:hard-decoding}. We compare the
raw gain from any canonical unilateral deviation at $\widehat q$
with its normalized quadratic gain.

Every canonical displacement of a shared quantity from its baseline
has magnitude at most $2\delta H$.  There are at most four firms in
a market.  Suppose every third partial derivative of
$q_i[-J(\sum_jq_j)]$ on $[0,2]^4$ has magnitude at most $2^{22}$.
Taylor's theorem bounds a unilateral payoff-difference remainder
in one shared market by at most
$2\cdot2^{22}(8\delta H)^3$.
Summing at most $72$ such remainders gives less than
$2^{39}\delta^3H^3$.
If $0\leq J(t)\leq32t^2$ for $0\leq t\leq1$, the two private
revenues contribute at most $2^{10}\delta^3H^3$ to a unilateral
difference.  The private revenue has no linear or quadratic term at
zero, whereas the private linear bias is exact.
The shared own-linear terms cancel by
\eqref{eq:relay-linear-cancellation}; opponent-only terms cancel
in a unilateral comparison.  Thus the total normalized error is
at most
\begin{equation}
                            2^{40}\delta H^3.
 \label{eq:relay-taylor-error}
\end{equation}

Choose the accuracy parameters to satisfy
\begin{equation}
 \frac{\varepsilon_*+1600L_0\rho}{\delta^2}
           +2^{40}\delta H^3<\xi,
 \qquad \sqrt{60k\,10^k\xi}<1/1000.
 \label{eq:relay-error-requirements}
\end{equation}
By \eqref{eq:decoded-raw-regret} and \eqref{eq:relay-taylor-error},
the first expression in \eqref{eq:relay-error-requirements} bounds
every quadratic unilateral gain at the decoded profile. Thus this
profile is a $\xi$-equilibrium of the quadratic game.
Lemma~\ref{lem:quadratic-source-regret} gives the claimed source
regret for the choice $\xi\leq1$ in Section~\ref{sec:constants}.

Finally add $\lambda\|q_i\|_2^2$ to every cost, where
$\lambda=\varepsilon_* /(4\cdot1000^2)$, and set
$\varepsilon_0=\varepsilon_*/2$.
The old regret of an $\varepsilon_0$-equilibrium in the perturbed
game is at most
$\varepsilon_0+\lambda1000^2=3\varepsilon_*/4<\varepsilon_*$.
The Hessian is now at least $2\lambda I$, and global cost
monotonicity and the zero-at-origin property are preserved.
Section~\ref{sec:constants} verifies all parameter inequalities
with fixed rational constants. Each directed payoff entry uses a
constant number of firms and markets. The cost coefficients are
obtained from the rational losses by a bounded number of additions
and multiplications per coefficient, and rational output profiles
are decoded by normalizing each private pair. The construction and
decoding therefore take polynomial time. The source hardness theorem
gives PPAD-hardness, and Appendix~\ref{sec:hard-membership} gives
membership for the stated bounded rational-output relations.
\end{proof}

\subsection{Fixed demands and relay parameters}
\label{sec:constants}

For each boundary pair $(r,h)$, we must fix the demand and all
reduction parameters independently of the source game. Start with a
demand that is nonincreasing and concave on the entire nonnegative
half-line and has enough curvature at output one to permit
amplification. Its curvature fixes the relay coefficients. We can
then choose the chain length for the required gain and the delivery
scale and penalty for the required accuracy. For the three degrees,
take $p_r(Q)=A-J_r(Q)$, where
\begin{align*}
 J_4(Q)&=\tfrac34Q^4-Q^3+\tfrac12Q^2,\\
 J_5(Q)&=\tfrac65Q^5-2Q^4+Q^3,\\
 J_6(Q)&=\tfrac{10}{3}Q^6-8Q^5+7Q^4-\tfrac83Q^3+\tfrac12Q^2.
\end{align*}
Their first derivatives vanish at zero and equal one at one; their
values at one are $J_r(1)=1/r$.  Their second derivatives are
\[
 J_4''(Q)=(3Q-1)^2,\quad
 J_5''(Q)=6Q(2Q-1)^2,\quad
 J_6''(Q)=(10Q^2-8Q+1)^2.
\]
They are nonnegative on the entire nonnegative half-line, so all
three final demands are decreasing and concave there.
Use
\[
 (s,r,\kappa)\in\{(1,6,9),(2,5,6),(3,4,4)\},\qquad
 \beta=1/10000,\qquad\vartheta=s/2.
\]
The relay coefficients are the following exact rationals, written
as terminating decimals:
\begin{center}
\begin{tabular}{c|rrrr}
$s$ & $A_0$ & $B_0$ & $D$ & $d_0$\\\hline
1 & 4.99955 & 0.50045 & 4.750675 & 4.750675\\
2 & 13.9976 & 2.0012 & 14.0012 & 12\\
3 & 22.4946 & 4.5018 & 25.5009 & 19.4985
\end{tabular}
\end{center}
These entries verify \eqref{eq:relay-required-bounds} by rational
comparison.  For the spectral gaps, use $A_0B_0<2.51,29,102$ and
$D>4.75,14,25$, respectively.  The first gap is greater than
$4.75-2(1.6)>3/2$; the other two exceed three.

To verify the gain bounds, let $\lambda_\pm$ be the two roots of
$x^2-Dx+A_0B_0$.  They are positive and
$\lambda_+-\lambda_->1$.  Evaluation at $(49/50)A_0>D/2$ shows
$\lambda_+<(49/50)A_0$: after division by $A_0$ the polynomial
value is
\[
       (2401/2500)A_0-(49/50)D+B_0>0.
\]
For the three rows, it suffices to use the rational bounds
$A_0>4.99$, $D<4.76$, $B_0>0.5$ for $s=1$;
$A_0>13.9$, $D<14.1$, $B_0>2$ for $s=2$; and
$A_0>22.4$, $D<25.6$, $B_0>4.5$ for $s=3$.
The determinant recurrence gives
\[
 \theta_k=
 \frac{\lambda_+^{k+1}-\lambda_-^{k+1}}{\lambda_+-\lambda_-},
 \qquad K>(50/49)^{k+1}.
\]
Set $k=2048$.  Since $(50/49)^{49}>2$, this implies $K>2^{41}$.
The weighted inverse bound applied to $T_ky=A_0e_1$ gives
$y_k<6\cdot2^k$ and therefore
$K=A_0y_k<2^{k+8}$.
Then $K_-=K(B_0/A_0)^{k+1}<1$.
The symmetric matrix similar to $T_k$ has smallest eigenvalue
greater than $3/2$, so each inverse diagonal is at most $2/3$;
these diagonal entries are unchanged by diagonal similarity.
Consequently $E<(2/3)\,102<100$.

With the chain length fixed, choose the remaining constants as follows.
\begin{equation}
 \begin{gathered}
 H=2^{k+4},\qquad \xi=2^{-30k-100},\qquad
 \delta=\frac{\xi}{2^{50}H^3},\qquad R=L_0=2^{100},\\
 \rho=\frac{\delta^2\xi}{10000L_0},\qquad
 M=\frac{3R}{\rho^2},\qquad
 \varepsilon_*=\frac{\delta^2\xi}{8},\\
 V=2^{90},\qquad A=V+10+8MH/\delta+1000M.
 \end{gathered}
 \label{eq:unified-hardness-constants}
\end{equation}
They give $2\delta H<\beta/2$, $\delta K<1$, and
\eqref{eq:shift-requirement}.  Since $60k<2^k$ and $10^k<2^{4k}$,
the relay error is less than $2^{3k}\sqrt\xi<1/1000$.
The other inequality in \eqref{eq:relay-error-requirements} follows
from the explicit budget
\[
 \frac18\xi+\frac{1600}{10000}\xi+\frac1{1024}\xi<\xi.
\]

To check the payoff and derivative bounds needed for decoding,
observe that on the capacity box all market totals are at most
$4000$.  The coefficient bounds
$J_r(Q)\leq32Q^6$ and $|J_r'(Q)|\leq100Q^5$ for $Q\geq1$ give
$J_r(Q)<V$ on this range; hence the final prices are positive.
They also bound each firm's raw payoff in absolute value by
$1000(32\cdot4000^6+5)<2^{100}$, and each relevant partial by
$32\cdot4000^6+100000\cdot4000^5+5<2^{100}$.
The same bounds for $Q<1$ follow by summing coefficient magnitudes.
For $0\leq t\leq1$, that sum gives $0\leq J_r(t)\leq32t^2$.
Finally, a third partial of $q_i[-J_r(Q)]$ has magnitude at most
$3|J_r''(Q)|+2|J_r'''(Q)|$ on $[0,2]^4$.
For $0\leq Q\leq8$, the displayed derivative formulas bound this
quantity by $2^{22}$ for every $r$.  For the largest case, $r=6$,
put $h(Q)=10Q^2-8Q+1$.  The bounds $|h(Q)|<2^{10}$ and
$|h'(Q)|<2^8$ give
$3|J_6''|+2|J_6'''|<3\cdot2^{20}+2\cdot2^{19}=2^{22}$;
the quartic and quintic bounds are smaller.
These choices satisfy the reduction bounds independently of the
size and payoffs of the source game. We can now combine the three
boundary families with the polynomial-time regimes.

\begin{proof}[Proof of Theorem~\ref{thm:classification}]
Proposition~\ref{prop:upper-algorithm} proves every P entry of
Table~\ref{tab:classification}. Theorem~\ref{thm:hardness} proves
the three boundary hard entries. Since degree and market participation
are upper bounds, each larger class contains the corresponding hard
instances. Appendix~\ref{sec:hard-membership} supplies membership for
the bounded-local-dimension search relations.
\end{proof}

\newcounter{cournotalgorithm}
\newenvironment{cournotpseudocode}[2]{%
  \begin{figure}[htbp]\centering\begin{minipage}{\linewidth}%
  \refstepcounter{cournotalgorithm}\label{#2}%
  \hrule\vspace{6pt}%
  \textbf{Algorithm \thecournotalgorithm.}\quad #1\par\smallskip
}{%
  \par\vspace{5pt}\hrule\end{minipage}\end{figure}%
}

\section{A marginal-charge algorithm for total-production costs}
\label{sec:total-output}

Now suppose that a firm pays $C_i(S_i)$ on its total output.
Shifting a unit between two markets leaves its production cost
unchanged. To exploit this in Theorem~\ref{thm:total-output}, give
firm $i$ a charge $\lambda_i$ per unit sold, the same on all its
markets. We can then solve each market separately for the quantities
firms would sell if they paid these charges. On the production side,
we ask each firm how much it would supply if it received $\lambda_i$
per unit. We seek charges at which each firm's total sales equal its
supply. At such charges the two payments cancel in the firm's profit.

Start with charges so high that sales are zero. Lower the charge of
a firm whose supply exceeds its sales. Its desired production can
fall and its sales can rise, displacing sales by its rivals.
For the update to make progress, the selected firm's gap must shrink
by more than all the rivals' gaps grow in total. We establish a bound
on this difference, use it to reduce the sum of the gaps geometrically,
and stop when the remaining gaps certify the requested profit regret
after rounding.

Use the input and preprocessing conventions of Section~\ref{sec:model}.
Inverse demand $p_{ij}$ may depend on the firm as well as the market.
Write $r$ for the largest demand degree and $B=\sum_i D_i$ for total
capacity.

\paragraph{Normalization of marginal costs.}
We may assume that every cost is nondecreasing.  Indeed, for
\[
 c=\max\{0,-\min_i C_i'(0)\},
\]
replace $p_{ij}(Q)$ by $p_{ij}(Q)+c$ and $C_i(s)$ by $C_i(s)+cs$.
Convexity makes the new costs nondecreasing, and every firm's
profit is unchanged.  Henceforth use the shifted functions and set
\[
 P=\max\{1,\max_{(i,j)\in E}p_{ij}(0)\}.
\]

\subsection{Market and supply responses at fixed charges}

Fix the charges and consider a trial total output $Q$ on one market.
For each active firm, equating marginal revenue to its charge determines
its delivery at that $Q$. We seek the value of $Q$ at which these
deliveries sum to $Q$ itself. To give this equation a unique solution
even for flat demand, and to make each firm's supply single-valued,
choose a positive rational $\tau$ and replace the demands and costs by
\begin{equation}
 \bar p_{ij}(Q)=p_{ij}(Q)-\tau Q,\qquad
 \bar C_i(s)=C_i(s)+\tau s^2.
 \label{eq:alg-regularization}
\end{equation}
Extend $\bar p_{ij}$ beyond $B_j$ by its tangent line at $B_j$.
The extension is continuously differentiable, concave, and has
slope at most $-\tau$.  It is used only by the auxiliary oracles;
it changes no feasible payoff.  On feasible profiles the
regularization changes each profit by at most $2\tau B^2$.
Consequently a regret guarantee transfers back to the original
game with loss at most $4\tau B^2$.

For charges $\lambda\in[0,P]^n$, put
\begin{equation}
 t_{ij}(Q)=-\bar p_{ij}'(Q),\qquad
 a_{ij}(Q,\lambda_i)
       =\frac{(\bar p_{ij}(Q)-\lambda_i)_+}{t_{ij}(Q)}.
 \label{eq:alg-responses}
\end{equation}
Here $t_{ij}\geq\tau$, and $t_{ij}$ is nondecreasing.  Market $j$
has an aggregate response $Q_j(\lambda)$ defined by
\begin{equation}
       \sum_{i:(i,j)\in E}a_{ij}(Q_j,\lambda_i)=Q_j,
       \qquad q_{ij}(\lambda)=a_{ij}(Q_j,\lambda_i).
 \label{eq:alg-market-root}
\end{equation}

\begin{lemma}[Scalar market response]
\label{lem:alg-market}
Equation~\eqref{eq:alg-market-root} has a unique nonnegative root
in $[0,P/\tau]$.  The response is continuous in $\lambda$, and
\begin{equation}
 \bar p_{ij}(Q_j)+q_{ij}\bar p'_{ij}(Q_j)=\lambda_i
       \quad(q_{ij}>0),\qquad
 \bar p_{ij}(Q_j)\leq\lambda_i\quad(q_{ij}=0).
 \label{eq:alg-marginal-revenue}
\end{equation}
At rational trial points, the scalar equation can be evaluated
exactly using rational arithmetic.
\end{lemma}
\begin{proof}
On an interval where $a_{ij}>0$,
\[
       \frac{\partial a_{ij}}{\partial Q}
                =-1-a_{ij}\frac{t'_{ij}}{t_{ij}}\leq-1.
\]
The function is zero otherwise.  Therefore
$\Psi_j(Q)=\sum_i a_{ij}(Q,\lambda_i)-Q$ is continuous and
strictly decreasing, with rate at least one.  It is nonnegative
at zero.  Since $\bar p_{ij}(Q)\leq P-\tau Q$, all its
positive-part terms vanish at $P/\tau$.  This proves existence
and uniqueness.  The same strict slope bound gives continuity
of the root in the charges; the remaining claims follow directly
from the definitions.
\end{proof}

The firm's supply response is
\begin{equation}
 \sigma_i(\lambda_i)=\mathop{\rm argmax}_{0\leq s\leq D_i}
                         \{\lambda_i s-\bar C_i(s)\}.
 \label{eq:alg-supply}
\end{equation}
It is single-valued, nondecreasing, and $1/(2\tau)$-Lipschitz.
It equals zero below $\bar C_i'(0)$, equals $D_i$ above
$\bar C_i'(D_i)$, and is otherwise the unique root of
$\bar C_i'(s)=\lambda_i$.  Thus it too is evaluated by scalar
bisection.  Define the residual
\begin{equation}
 F_i(\lambda)=\sigma_i(\lambda_i)-S_i(\lambda),
       \qquad S_i(\lambda)=\sum_jq_{ij}(\lambda).
 \label{eq:alg-residual}
\end{equation}
Our invariant will be $F\geq0$.  Under this invariant the
auxiliary quantities are already feasible, since
$S_i\leq\sigma_i\leq D_i$.  At $\lambda=P\mathbf1$ every
market response is zero, so the invariant holds initially.

\subsection{Relative column dominance of the residual map}

Suppose we lower only $\lambda_i$. We need a quantitative comparison
between the decrease in firm $i$'s residual and the increases in its
rivals' residuals. Read these effects down column $i$ of the Jacobian
of $F$. The diagonal entry measures the response of firm $i$ and the
other entries measure the responses of its rivals. We first compare
them while the active firms stay fixed, then extend the bound across
changes in market participation.

\begin{lemma}[Relative column dominance]
\label{lem:alg-columns}
Suppose
\begin{equation}
       \frac{Q t'_{ij}(Q)}{t_{ij}(Q)}\leq K
       \qquad(Q>0)
 \label{eq:alg-curvature}
\end{equation}
on every continued demand curve.  Put $\gamma=1/(n+K)$.
Along every coordinate segment,
\begin{equation}
 \partial_iF_i\geq0,\qquad \partial_iF_k\leq0\ (k\ne i),
 \qquad
 \sum_{k\ne i}(-\partial_iF_k)
                    \leq(1-\gamma)\partial_iF_i.
 \label{eq:alg-column}
\end{equation}
These inequalities are understood on smooth portions and remain
valid after integration.  Also $F_i$ is $(m+1)/\tau$-Lipschitz
in its own charge.
\end{lemma}
\begin{proof}
Fix one market with positive aggregate $Q$ and active firms $A$.
For $i\in A$, write
\[
 h_i=1/t_i(Q),\quad
 w_i=1+q_i t_i'(Q)/t_i(Q),\quad Z=1+\sum_{i\in A}w_i.
\]
Differentiating its scalar equation with respect to an active
firm's charge $\lambda_\ell$ gives
\begin{equation}
 \partial_{\lambda_\ell}Q=-h_\ell/Z,
 \qquad
 \partial_{\lambda_\ell}q_i
       =-h_i\mathbf1_{\{i=\ell\}}+w_i h_\ell/Z.
 \label{eq:alg-market-jacobian}
\end{equation}
An inactive charge has derivative zero.  In column $\ell$ of
the derivative of $-q$, the diagonal is
$h_\ell(Z-w_\ell)/Z$, the sum of the absolute off-diagonal
entries is $h_\ell(Z-1-w_\ell)/Z$, and their difference is
$h_\ell/Z$.  Moreover,
\[
 Z-w_\ell
 =|A|+\sum_{i\in A\setminus\{\ell\}}
       \frac{Qt_i'(Q)}{t_i(Q)}\frac{q_i}{Q}
 \leq n+K.
\]
This proves the relative inequality for one market.  Summing
over markets and adding the nonnegative supply derivative proves
\eqref{eq:alg-column}.  Every market contributes at most
$1/\tau$ to the diagonal; supply contributes at most
$1/(2\tau)$.

To justify integration across changes of active set, decrease one
charge while fixing the others. At the old aggregate the sum of the
firms' responses in~\eqref{eq:alg-market-root} weakly increases, so
the new market root is weakly larger. Each rival's response at its
fixed charge is nonincreasing in that root. The selected firm's quantity, equal to
the root minus the sum of rival quantities, is therefore nondecreasing.
A rival can exit at most once and the selected firm can enter at most
once. The aggregate can pass the tangent-extension boundary at most
once, and the supply inverse has only two clipping thresholds.
Integrating on these pieces and using continuity at their endpoints
gives, for $\lambda'$ obtained by lowering $\lambda_i$,
\[
 0\leq\sum_{k\ne i}\bigl(F_k(\lambda')-F_k(\lambda)\bigr)
 \leq (1-\gamma)\bigl(F_i(\lambda)-F_i(\lambda')\bigr).
\]
\end{proof}

For global polynomials, $K$ has a bound independent of coefficient
magnitudes.  In fact, we may take
\begin{equation}
                         K=2\max\{r-1,0\}^2.
 \label{eq:alg-K}
\end{equation}
To see this, $g=-\bar p_{ij}'$ is a positive nondecreasing
polynomial of degree at most $d=\max\{r-1,0\}$ on $[0,B_j]$.
On $[0,Q]$, its maximum is $g(Q)$.  Markov's polynomial
inequality~\cite{BE95} therefore gives
$Qg'(Q)\leq2d^2g(Q)$.  Beyond $B_j$ the slope is constant.
Thus the fraction $\gamma$ in the update bound depends only on $n$
and the demand degree.

\subsection{Rational evaluation of responses and residuals}

Even at rational charges the market root can be irrational. To
return rational quantities within capacity, we will round each
delivery down. Bisect the market equation and use the upper endpoint
$U\geq Q_j(\lambda)$ of the bracket. Since each firm's delivery is
nonincreasing in total output, $a_{ij}(U,\lambda_i)\leq q_{ij}(\lambda)$.
To choose the bracket width, bound the sum of the delivery errors.
For every
$Q\geq Q_j(\lambda)$,
\[
 \sum_i a_{ij}(Q,\lambda_i)\leq Q_j(\lambda)\leq Q.
\]
Consequently, on smooth intervals,
\[
 -\frac{d}{dQ}\sum_i a_{ij}(Q,\lambda_i)
 \leq n+K\frac{\sum_i a_{ij}(Q,\lambda_i)}{Q}
 \leq n+K.
\]
Integration proves
\begin{equation}
 \sum_i\bigl(q_{ij}(\lambda)-a_{ij}(U,\lambda_i)\bigr)
          \leq(n+K)(U-Q_j(\lambda)).
 \label{eq:alg-lower-error}
\end{equation}
If the root is zero then all responses vanish, and the same
conclusion holds directly.

\begin{cournotpseudocode}{\textsc{LowerResponse}$(\lambda,\delta)$}{alg:response}
\emph{Input:} rational $\lambda\in[0,P]^n$ and $\delta>0$.
\begin{enumerate}[label=\arabic*.,leftmargin=2em,itemsep=2pt,topsep=4pt]
\item For each market $j$, perform the following steps.
  \begin{enumerate}[label=(\alph*),leftmargin=2em,itemsep=2pt,topsep=2pt]
  \item Set $L=0$ and $U=P/\tau$.
  \item While $U-L>\delta/[m(n+K)]$, set $x=(L+U)/2$ and
        evaluate $z=\sum_{i:(i,j)\in E}a_{ij}(x,\lambda_i)-x$.
        If $z>0$, set $L=x$; otherwise set $U=x$.
  \item Set $y_{ij}=a_{ij}(U,\lambda_i)$ for every incident firm $i$.
  \end{enumerate}
\item Return $y$.
\end{enumerate}
\end{cournotpseudocode}

Algorithm~\ref{alg:response} returns rational $y$ satisfying
\begin{equation}
             0\leq y\leq q(\lambda),\qquad
             \|q(\lambda)-y\|_1\leq\delta.
 \label{eq:alg-response-guarantee}
\end{equation}
This holds even when the auxiliary response is infeasible.
The tangent extension in~\eqref{eq:alg-regularization} is therefore
used during trial evaluations.

To implement \textsc{Residual}$(\lambda,\nu)$, first compute
$y=\textsc{LowerResponse}(\lambda,\nu/2)$.  For each firm,
compute a supply estimate $\widetilde\sigma_i$ to error at most
$\nu/2$.  The clipping cases in~\eqref{eq:alg-supply} are tested
exactly; in the interior, bisect $[0,D_i]$ until its width is at
most $\nu$ and return its midpoint.  Set
\[
                   v_i=\widetilde\sigma_i-\sum_j y_{ij}.
\]
Then $|v_i-F_i(\lambda)|\leq\nu$ for every firm.  All signs
used by both bisections are signs of explicit rational numbers.

\subsection{Feasibility and contraction of charge updates}

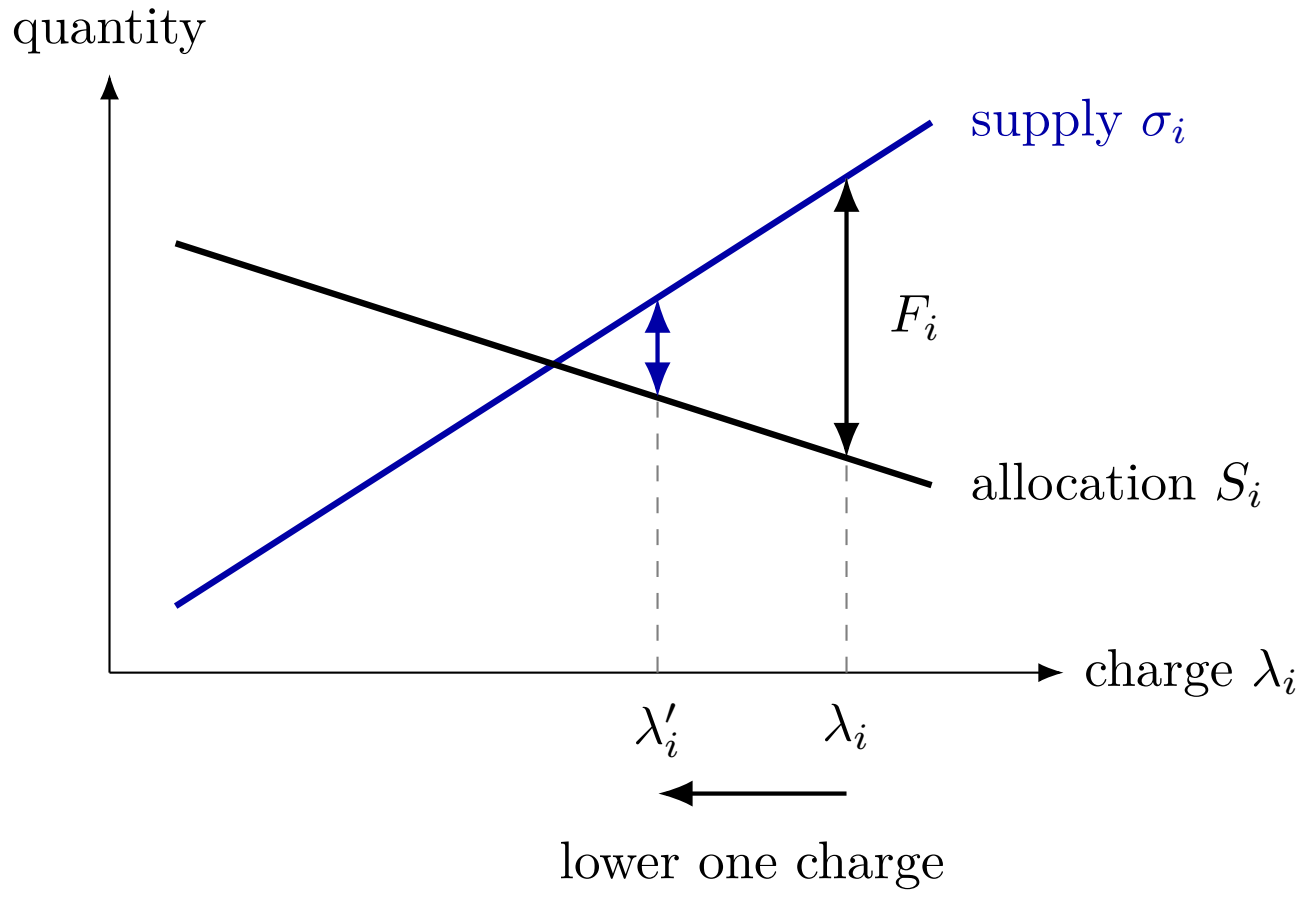
\begin{figure}[t]
\centering
\begin{tikzpicture}[x=1.25cm,y=.8cm,>=Latex,font=\small]
 \draw[->] (0,0)--(5.05,0) node[right] {charge $\lambda_i$};
 \draw[->] (0,0)--(0,4.95) node[above] {quantity};
 \draw[very thick,blue!65!black] (.35,.55)--(4.35,4.55);
 \node[anchor=west,blue!65!black] at (4.45,4.55) {supply $\sigma_i$};
 \draw[very thick] (.35,3.55)--(4.35,1.55);
 \node[anchor=west] at (4.45,1.55) {allocation $S_i$};
 \draw[dashed,gray] (3.9,0)--(3.9,4.1);
 \draw[dashed,gray] (2.9,0)--(2.9,3.1);
 \draw[<->,thick] (3.9,1.775)--node[right=4pt] {$F_i$}(3.9,4.1);
 \draw[<->,thick,blue!65!black] (2.9,2.275)--(2.9,3.1);
 \draw[->,thick] (3.9,-1.0)--node[below=5pt]{lower one charge}(2.9,-1.0);
 \node[below=2pt] at (3.9,0) {$\lambda_i$};
 \node[below=2pt] at (2.9,0) {$\lambda_i'$};
\end{tikzpicture}
\caption{One coordinate update, with other charges fixed. The curves
are schematic. Lowering $\lambda_i$ brings the firm's supply and
allocated quantity closer while keeping their difference nonnegative.
The column inequality bounds the resulting increase in other firms'
residuals.}
\label{fig:charges}
\end{figure}

Use the residual estimates to choose a firm with a largest gap.
We lower its charge until the estimated gap is roughly half its
old value (Figure~\ref{fig:charges}). Aiming for half leaves room
for estimation error while keeping the true residual positive.
The other firms' residuals can only grow during this update, so
every firm stays within capacity. To bound the number of updates,
we use Lemma~\ref{lem:alg-columns} to limit how much of the selected
decrease those increases can offset.

Let $a>0$ be the desired residual bound and put $\nu=a/33$.
In Algorithm~\ref{alg:descend}, $\lambda^{(i\leftarrow x)}$ denotes
$\lambda$ with coordinate $i$ replaced by $x$.

\begin{cournotpseudocode}{Descending marginal charges}{alg:descend}
\emph{Input:} regularization $\tau>0$, residual target $a>0$,
and output accuracy $\eta>0$.
\begin{enumerate}[label=\arabic*.,leftmargin=2em,itemsep=2pt,topsep=4pt]
\item Set $\lambda=P\mathbf1$ and $\nu=a/33$.
\item Compute $v=\textsc{Residual}(\lambda,\nu)$ and choose
      $i\in\arg\max_k v_k$.
\item If $v_i\leq32\nu$, return
      $\textsc{LowerResponse}(\lambda,\eta)$.
\item Set $v_*=v_i$, $L=0$, and $U=\lambda_i$.
\item Set $x=(L+U)/2$ and compute
      $w=\textsc{Residual}(\lambda^{(i\leftarrow x)},\nu)_i$.
\item If $|w-v_*/2|\leq v_*/16$, set $\lambda_i=x$ and return to Step~2.
\item If $w>v_*/2$, set $U=x$; otherwise set $L=x$.
      Return to Step~5.
\end{enumerate}
\end{cournotpseudocode}

\begin{lemma}[Feasibility and contraction]
\label{lem:alg-contraction}
Every accepted iterate satisfies $F\geq0$.  At termination,
$\max_iF_i\leq a$.  If $R=\sum_iF_i$, each nonterminal update
satisfies
\begin{equation}
               R_{\rm new}\leq
                 \left(1-\frac{1}{3n(n+K)}\right)R.
 \label{eq:alg-contraction}
\end{equation}
There are
$O(n(n+K)(1+\log_+(B/a)))$ accepted updates, where
$\log_+x=\max\{0,\log x\}$.
\end{lemma}
\begin{proof}
The initial invariant was established above.  In a nonterminal
iteration $v_*>32\nu$.  The selected old residual is at least
$v_*-\nu>31v_*/32$, whereas setting its charge to zero gives
$F_i\leq0$, since $\sigma_i(0)=0$.  Thus the target $v_*/2$
is bracketed.  A failed acceptance test has the correct sign:
$v_*/16>\nu$.

At acceptance, the new true residual lies in
$[13v_*/32,19v_*/32]\subseteq[3v_*/8,5v_*/8]$.  It is positive;
all other residuals weakly increase when $\lambda_i$ decreases.
The entire accepted coordinate segment therefore keeps $F\geq0$.
Let $\Delta=F_i^{\rm old}-F_i^{\rm new}$.  Every old residual
is at most $v_*+\nu<33v_*/32$, so
\[
 \Delta\geq\frac{11v_*}{32}\geq\frac{R}{3n}.
\]
The integrated column inequality bounds the total increase of
all other residuals by $(1-\gamma)\Delta$.  Thus the total
residual decreases by at least $\gamma\Delta$, proving
\eqref{eq:alg-contraction}.

The selected-coordinate Lipschitz bound is
$\mathcal M=(m+1)/\tau$.  Once the inner interval has width at
most $v_*/(32\mathcal M)$, its midpoint has true target error
at most $v_*/64$.  Adding the oracle error $\nu<v_*/32$ makes
the acceptance test succeed.  Hence every inner search terminates.

At stopping, $F_k\leq v_i+\nu\leq33\nu=a$.  Initially
$R\leq\sum_iD_i=B$, and the stopping test must succeed once
$R\leq31\nu$.  The iteration bound follows from the contraction.
\end{proof}

\subsection{Profit regret after rounding}

The stopping test controls a quantity gap. We need a bound on profit
from any unilateral deviation, including one that changes total output.
At the exact market response the firm already maximizes revenue less
its charge against fixed rivals. We can therefore bound its gain by
comparing two total outputs at that charge. We then account for the
downward rounding of both its own deliveries and its rivals' deliveries.

\begin{lemma}[A one-dimensional regret certificate]
\label{lem:alg-regret}
At any charge vector with $F\geq0$, its exact auxiliary response
is feasible and has regularized-game regret at most $PF_i$ for
firm $i$.
\end{lemma}
\begin{proof}
For fixed opponents, the function
\[
        z_i\longmapsto
        \sum_jz_{ij}\bar p_{ij}(Q_{-i,j}+z_{ij})
                   -\lambda_i\sum_jz_{ij}
\]
is concave.  By~\eqref{eq:alg-marginal-revenue}, its gradient
at $q_i$ is zero on positive coordinates and nonpositive on zero
coordinates.  Thus $q_i$ maximizes this function over feasible
deviations.  Restoring the production cost bounds the gain from
any deviation by
\[
 [\lambda_i\sigma_i-\bar C_i(\sigma_i)]
       -[\lambda_iS_i-\bar C_i(S_i)].
\]
Since $S_i\leq\sigma_i$, nondecreasing cost bounds this difference
by $\lambda_i(\sigma_i-S_i)\leq PF_i$.
\end{proof}

\begin{lemma}[Downward rounding]
\label{lem:alg-round}
For any feasible $0\leq y\leq q$, nonincreasing demands bounded
above by $P$ and nondecreasing total costs satisfy
\begin{equation}
              \operatorname{Reg}_i(y)
              \leq\operatorname{Reg}_i(q)+P\|q-y\|_1.
 \label{eq:alg-downward}
\end{equation}
\end{lemma}
\begin{proof}
Fix a deviation $z_i$ against $y_{-i}$ and put
$\delta_j=Q_{-i,j}(q)-Q_{-i,j}(y)\geq0$ and
$z'_{ij}=(z_{ij}-\delta_j)_+$.  This is a feasible deviation
against $q_{-i}$.  When $z_{ij}\geq\delta_j$, the two deviations
face the same aggregate and their revenues differ by at most
$P\delta_j$.  When $z_{ij}<\delta_j$, the comparison deviation
is zero and the same upper bound holds.  Its cost is no larger.
Hence the best-response value against $y_{-i}$ exceeds that
against $q_{-i}$ by at most $P\sum_j\delta_j$.

For the current payoffs, decreasing quantities raises market
prices and lowers production cost.  Consequently
$\pi_i(q)-\pi_i(y)\leq P\sum_j(q_{ij}-y_{ij})$.
Adding the two inequalities proves~\eqref{eq:alg-downward}.
The argument also permits negative prices.
\end{proof}

The returned $y$ is componentwise below the feasible response at
termination and is therefore feasible.  Combining the last two
lemmas,~\eqref{eq:alg-response-guarantee}, and regularization gives
\begin{equation}
       \operatorname{Reg}_i(y)
          \leq Pa+P\eta+4\tau B^2.
 \label{eq:alg-error-budget}
\end{equation}
For a target $0<\varepsilon\leq1$, choose the precisions so that
each of the three terms in~\eqref{eq:alg-error-budget} is at most
$\varepsilon/4$. Set
\begin{equation}
 \boxed{\qquad
 \tau=\frac{\varepsilon}{16(1+B)^2},\quad
 a=\eta=\frac{\varepsilon}{4P},\quad
 \nu=\frac a{33},\quad K=2\max\{r-1,0\}^2.
 \qquad}
 \label{eq:alg-parameters}
\end{equation}
Then $4\tau B^2\leq\varepsilon/4$ and
$P(a+\eta)=\varepsilon/2$.  Equation~\eqref{eq:alg-error-budget}
therefore gives regret at most $3\varepsilon/4<\varepsilon$.

\subsection{Running time}
\label{sec:alg-running-time}

Each charge update calls a bisection that in turn evaluates market
and supply responses. We count these nested calls, then bound the
bit length of the rational numbers used in them, with the precisions
chosen in~\eqref{eq:alg-parameters}.

Store the current charge vector and the residual estimates in arrays.
At a midpoint of the inner search, evaluate the trial residual once
and use the stored value for both the acceptance and sign tests.
On acceptance, update the selected charge in the stored vector.
Thus reading a later charge does not repeat the searches that produced
earlier iterates. A market-root
bisection with aggregate quantity error at most $\delta$ uses
\[
 O\!\left(1+\log_+
      \frac{m(n+K)P}{\tau\delta}\right)
\]
iterations.  A supply bisection to error $\delta$ uses
$O(1+\log_+(B/\delta))$ iterations.  Since every nonterminal
selected estimate exceeds $32\nu$, an inner coordinate search
uses
\[
 O\!\left(1+\log_+
             \frac{P(m+1)}{\tau a}\right)
\]
residual evaluations.  Lemma~\ref{lem:alg-contraction} bounds the
accepted updates by
\begin{equation}
 O\!\left(n(n+r^2)
       \left(1+\log_+\frac{PB}{\varepsilon}\right)\right).
 \label{eq:alg-update-count}
\end{equation}
Each oracle iteration evaluates explicitly represented
polynomials and their derivatives, so its number of arithmetic
operations is polynomial in the total coefficient count.

These operation bounds also give the claimed standard running
time.  Every updated charge is $P$ times a dyadic rational; its
denominator exponent increases by at most the number of inner
bisections.  Market trial points are dyadic multiples of
$P/\tau$, and supply trial points are dyadic multiples of $D_i$.
All their encodings consequently have polynomial length.
Exact polynomial evaluations and sums of polynomially many
rational terms preserve this bound.  Evaluated values are used
for sign tests and output, not as new unconstrained endpoints in
the nested bisections.  Finally, all quantities
in~\eqref{eq:alg-parameters} have encoding length polynomial in
the input length and $b$.

\begin{proof}[Proof of Theorem~\ref{thm:total-output}]
Take $\varepsilon=2^{-b}$ and choose the parameters
in~\eqref{eq:alg-parameters}. Lemma~\ref{lem:alg-contraction}
gives a feasible terminal response with $\max_i F_i\leq a$.
Algorithm~\ref{alg:response} returns a feasible rational profile
below this response with total rounding error at most $\eta$.
Lemmas~\ref{lem:alg-regret} and~\ref{lem:alg-round}, followed by
the regularization bound, give original-game regret at most
$P(a+\eta)+4\tau B^2\leq3\varepsilon/4$.
The operation and bit bounds above are polynomial in $\inputsize$
and $b$. Restoring the forced zero deliveries completes the construction.
\end{proof}

\section{Open questions}

Which ways of coupling a firm's deliveries through its cost preserve
tractability? A cost of total production permits high-precision
approximation at every demand degree. A general convex quadratic cost
can already give hardness at degree four when four firms share a market.
Between these cases, can one characterize a larger family of allocation
costs that still admits efficient equilibrium computation?

Even for costs of total production, the demand representation and the
choice of adjustment rule raise distinct questions. Our running-time
bound uses a polynomial on the entire feasible interval to control
curvature. Can one obtain a comparable bound for succinctly represented
piecewise polynomial demands? Our algorithm adjusts auxiliary
marginal charges. How quickly can natural adjustment rules converge
when firms update their quantities directly?

\paragraph{AI Use Statement.}
This work involved extensive, author-directed exchanges with OpenAI
Astra, including suggestions and critical evaluation of its responses
and the proposed ideas and arguments. OpenAI Astra also assisted in
writing and revising the manuscript in response to detailed prompts.
Gemini Argon assisted with parts of the work.

\bibliographystyle{abbrvnat}
\bibliography{network_cournot_complexity}

\appendix
\section{PPAD membership at fixed degree and local dimension}

\label{sec:hard-membership}

To place the bounded search problem in PPAD, we need a short rational
delivery profile whose regret bound holds against every feasible real
deviation. We first look for an approximate fixed point of a projected
marginal-profit map and choose its accuracy so that the regret bound
survives rounding. We then use the bounds on degree and local dimension
to check the output by testing, one firm at a time, whether any
deviation gains more than the regret tolerance.

Fix constants
$k_0,r_0$.  A firm has at most $k_0$ quantity coordinates, demands have
degree at most $r_0$, costs are convex quadratic polynomials, and
capacities and coefficients are rational.  The shape conditions are
input promises.  In particular, our hard classes have $k_0\leq74$
and $r_0\leq6$.

Let $d$ be the total number of quantity coordinates, remove zero-capacity
coordinates, and write $X=\prod_iX_i$ for the product of capacity
simplexes.  The case $d=0$ is trivial; assume $d\geq1$.  Define
\[
 g_i(q)=\nabla_{q_i}\pi_i(q),\qquad
 F_i(q)=\operatorname{proj}_{X_i}(q_i+g_i(q)).
\]
Projection is rational and polynomial-time computable: take the positive
part if its sum is within capacity; otherwise sort and subtract the
unique common threshold whose positive parts sum to capacity.
Coefficient bounds give rational $D,G,H\geq1$ with $D\geq\max_iD_i$,
$|g_{ij}|\leq G$, and every derivative of $g$ bounded by $H$ on $X$.
Their encoding lengths are polynomial in the input length, and
$F$ is Lipschitz in the maximum norm with constant
$\Lambda=d(1+dH)$.
The polynomially computable and polynomially continuous fixed-point
framework of Etessami and Yannakakis \cite[Proposition~2(2)]{EY10}
therefore computes a weak approximate fixed point through a PPAD
reduction.

To obtain actual regret, put $r_i=F_i(q)-q_i$.  The projection
inequality and concavity of own profit give, for every $z_i\in X_i$,
\[
 \pi_i(z_i,q_{-i})-\pi_i(q)
 \leq g_i(q)^\top(z_i-q_i)
 \leq \|r_i\|_2(2D_i+\|g_i(q)\|_2).
\]
Thus regret is at most $d(2D+dG)\|F(q)-q\|_\infty$.
Let
\[
 \eta=\frac{\min\{1,\varepsilon\}}{d(2D+dG)},\qquad
 2^{-t}\leq\frac{\eta}{2(1+\Lambda)},
\]
with $t$ the least nonnegative integer satisfying the inequality.
Request residual at most $\eta/2$ and round every coordinate down
to a multiple of $2^{-t}$.  Feasibility is preserved and the new
residual is at most $\eta$.  We restrict outputs to rational profiles
of total encoding length at most
\[
               d\bigl(2t+\lceil\log_2(1+D)\rceil+10\bigr).
\]
This is an explicit polynomial bound.

For such an output $q$, a violation of firm $i$'s regret condition is
the existence of $z_i\in X_i$ for which
$\pi_i(z_i,q_{-i})-\pi_i(q)>\varepsilon$.
There are at most $k_0$ variables, one for each market accessible to
firm $i$. Revenue has degree at most $r_0+1$ and cost has degree at
most two, so the violation polynomial has degree at most
$\max\{r_0+1,2\}$. The quadratic term can matter even under constant
inverse demand.
Fixed-dimensional real-algebraic decision \cite{BPR96} verifies this
condition in polynomial time.  The shape promises are also decidable
in polynomial time in this fixed-degree, fixed-local-dimension class;
invalid inputs may be assigned a distinguished trivial output.
The relation is consequently total,
polynomially balanced, and polynomial-time verifiable, and the weak
fixed-point reduction places it in PPAD.
The bounded local dimension is used for exact verification of the
regret inequality; it is unnecessary for the fixed-point map itself.

\end{document}